# Magnetic Field Inspired Contact Angle Hysteresis Drives Floating Polyolefin Rafts

Mark Frenkel[1], Victor Danchuk[2], Victor Multanen[1], Edward Bormashenko[1]

[1]*Ariel University, Engineering Faculty, Chemical Engineering, Biotechnology and Materials Department, P.O.B. 3, 407000, Ariel, Israel*

[2]*Ariel University, Exact Sciences Faculty, Physics Department, P.O.B. 3, 407000, Ariel, Israel*

Corresponding author: Edward Bormashenko

E-Mail: Edward@ariel.ac.il

Phone: 972-3-074 7296863

**ABSTRACT**

Displacement of floating polymer (polyolefin) rafts by steady magnetic field is reported. The effect is due to the interplay of gravity deforming the water/vapor interface, contact angle hysteresis and diamagnetic properties of the liquid support. Magnetic field (*ca.* 0.06 T) deformed the water/vapor interface and impacted the interfacial apparent contact angle. This deformation gave rise to the propulsion force, displacing the polymer raft. The velocity of displacement of center of mass of rafts $v_{cm} \cong 0.015 \frac{m}{s}$ was registered. The effect is related to the contact angle hysteresis affected by the magnetic field, enabling the change in the apparent interfacial contact angle. The semi-quantitative model of the process is suggested.

Contact angle hysteresis, exerted to intensive research recently, is responsible for a diversity of interfacial phenomena, including: equilibrium of a liquid column, trapped in a vertical capillary tube [1], equilibrium and sliding of droplets on inclined planes and curved surfaces [2-4], adhesion of droplets [5], stability of Cassie-like wetting on superhydrophobic surfaces [6] and last but not least for floating of micro-boats [7]. Contact angle hysteresis, which is the difference between the maximum (advancing) and minimum (receding) contact angle, inherent for a certain triad: solid/liquid/vapor is partially caused by adhesion hysteresis in the solid–liquid contact area (2D effect [8-11]) and partially by pinning of the solid–liquid-air triple line due to the surface roughness (1D effect [12-17]). The physical origin of the contact angle hysteresis remains debatable and it is reasonable to suggest, that it arises from a complicated interplay of chemical and physical sources [9, 18-20].

It is noteworthy, that the contact angle hysteresis usually retains the motion of droplets or solid objects [2-4, 13]. In contrast, we report surprising, paradoxical, experimental findings describing the situation, where contact angle hysteresis promotes the motion of floating polymer (polyolefin) rafts. In our experiments circular polyethylene (PE) and polypropylene (PP) rafts were placed on the water/vapor interface, as shown in **Figure 1**. De-ionized water was prepared from a synergy UV water purification system from Millipore SAS (France). The specific resistivity of water was $\hat{\rho} = 18.2 M\Omega \times cm$ at 25 °C. The diameter of polymer rafts were 5.6±0.1 mm with the thickness of 0.6±0.05mm and the mass of 0.0134±0.0001g for the PE rafts and the thickness 0.9±0.05mm and the mass of 0.0172±0.0001g for PP ones. The density of rafts was $\rho_{PP} = 0.95 \times 10^3 \text{kg/m}^3$; $\rho_{PE} = 0.92 \times 10^3 \text{kg/m}^3$. The rafts were cut from polyolefin films, manufactured by hot pressing. The advancing apparent contact angles for rafts were $\theta_{ad}^{PP} = 102.2 \pm 0.5°$; $\theta_{ad}^{PE} = 122.3 \pm 0.5°$, as established with the Rame-Hart 500 goniometer by the sessile droplet method. Thus, the rafts were hydrophobic. The motion of rafts was registered from above with a rapid camera (Casio EX-FH20). The experiments were carried out under ambient conditions.

Cylindrical Neodymium permanent magnets producing magnetic fields $B \cong 0.1 \pm 0.05 T$ were placed above the water/vapor interface, as depicted in **Figure 1**. The distance $h$ between the magnet and the water level was $h = 1.0 \pm 0.1 mm$; the maximal lateral distance between the magnet and the margin of the raft was $L=2\pm0.1$ cm (see **Figure 1**). In a contra-intuitive, paradoxical way, the rafts moved with a characteristic velocity of the center of mass $v_{cm} \cong 0.015 \text{m/s}$, as shown in **Figure 2** (see also the Supplementary Materials, Video 1). When the magnet was moved away it attracted polymer rafts (both PP and PE). Why do the observed displacements of rafts are surprising? Indeed, PE and PP are pronounced diamagnetic materials with the specific magnetic susceptibility of $\chi \cong -1.01 \times 10^{-8}$ m$^3$/kg [21-23]. Thus, the expected interaction of the magnet with the aft is negligible; moreover, magnetic fields push out diamagnetic materials [23]. In contrast, in our experiments the attraction of rafts to the permanent magnet was observed.

All the observed effect is due to the interplay of gravity, contact angle hysteresis and the diamagnetic properties of water ($\chi_{water} \cong -9.1 \times 10^{-9}$ m$^3$/kg). When the magnetic field acts on water, the water/vapor interface it is deformed, as depicted in **Figure 1** (see also the Supplementary Materials, Video 2); consequently the apparent interfacial contact angle is changed from α to β. This change gives rise to the horizontal force on the order of magnitude of $ca. \gamma l(\cos\beta - \cos\alpha)$, moving the raft, where $l$ is the characteristic lateral dimension of the raft (i.e. its radius in our case), and γ is the water/air surface tension. The similar strategy of actuation of droplets was reported recently in Ref. 24, where diamagnetic liquid marbles [25] were displaced by the steady magnetic field of *ca.* 50 mT. In turn, we demonstrate the same idea may be exploited for the actuation of the motion of floating solid polymer rafts.

The change is the apparent contact angle arising from the action of the magnetic field is illustrated with **Figure 3**. The menisci and interfacial apparent contact angles were visualized with the goniometer Rame-Hart 500. Note, that the Young contact angles are insensitive to external fields (gravitational, electric and magnetic) as demonstrated in Refs. 10, 26-27; contrastingly, apparent contact angles are affected by these fields. Hence, the reported effect is at least partially due to the contact angle hysteresis.[1,10] Note also, that the interfacial angles should not be mixed with the apparent and receding contact angles denoted by $\theta$ and $\theta_R$ [26-28]. The interrelation between contact angles is obvious: $\alpha = \theta - \frac{\pi}{2}$ and $\beta = \theta_R - \frac{\pi}{2}$. It is reasonable to suggest that the contact angle hysteresis in the described experiments origins in the pinning of the triple (three phase) line to the circumference of rafts [10, 13].

Why do we recognize the reported effect as surprising? At the first glance, it seems that extremely weak magnetic properties of involved media will demand very high magnetic fields for displacing the polymer boat. It turns out, that moderate fields on the order of magnitude of 0.1 T will be sufficient for performing this task. Consider a spherical water droplet of radius $R$ placed into the magnetic field $B$. Equating the surface energy of the droplet to the magnetic one yields:

$$\gamma 4\pi R^2 = \frac{B^2}{2\mu\mu_0} \frac{4}{3}\pi R^3, \qquad (1)$$

where $\mu$ is the magnetic permeability of water. Assuming $\mu \cong 1$ (water is a weak diamagnetic material, $\mu = 1 + \chi_m$, where $\chi_m$ is the molar magnetic susceptibility of water, $|\chi_m| \ll 1$), and omitting the numerical coefficients, gives rise to the dimensionless number, denoted by $\psi$, relating the effects owing to the surface tension to those due to the applied magnetic fields, supplied by Eq. 2:

$$\psi = \frac{B^2 R}{\mu_0 \gamma} \qquad (2)$$

It is seen from Eq. 2, that the effects due to the applied magnetic field became comparable when the value of the magnetic fields is on the order of magnitude of the value $B^*$, given by Eq. 3:

$$B^* \cong \sqrt{\frac{\mu_0 \gamma}{R}} \qquad (3)$$

It is plausible to assume that the value of $R$ is close to the so-called capillary length inherent for water, labeled $l_{ca}$ (see Refs. 1, 9-10). Thus, finally we derive:

$$\psi = \frac{B^2 l_{ca}}{\mu_0 \gamma}; \qquad (4a)$$

$$B^* = \sqrt{\frac{\mu_0 \gamma}{l_{ca}}} \qquad (4b)$$

Substituting $\mu_0 = 4\pi \times 10^{-7} \, \text{N/A}^2$; ; $\gamma \cong 70 \times 10^{-3} \, \text{N/m}$ and $l_{ca} \cong 2.7 \times 10^{-3} \, \text{m}$ into Eq. 4b yields the rough estimation: $B^* \cong 5 \times 10^{-3} \, \text{T}$. Indeed, Baigl et al. reported in Ref 24 actuation of liquid marbles supported by water by the magnetic fields of the order of magnitude of 50 mT, which is not far from the crude estimation supplied by Eq. 4. It is seen from Eqs. 4a-b, that the reported effect arises from the conjunction of gravity, interfacial and magnetic effects.

Now estimate the velocities of polymer rafts arising from the contact angle hysteresis, inspired by the external magnetic fields. Equating the driving force originating from the magnetically induced contact angle hysteresis to the viscous force and omitting the numerical factors yields the following crude estimation, applicable for the steady motion of the rafts:

$$\gamma(\cos\beta - \cos\alpha) \approx \eta l v_{cm}, \qquad (5)$$

where $\eta$ is the dynamic viscosity of water. Considering: $\cos\beta - \cos\alpha = 2\sin\frac{\beta+\alpha}{2} \times \sin\frac{\alpha-\beta}{2}; \frac{\alpha-\beta}{2} \ll 1$ and denoting $\alpha - \beta = \Delta\alpha$ yields:

$$v_{cm} \cong \sin\frac{\alpha+\beta}{2} \frac{\gamma}{\eta} \Delta\alpha \cong \sin\alpha \frac{\gamma}{\eta} \Delta\alpha. \qquad (6)$$

Substituting $\alpha \cong \beta = \frac{\pi}{6} \, rad; \Delta\alpha = 1.7 \times 10^{-2} \, rad; \eta = 9 \times 10^{-4} \, Pa \times s$ into Eq. 6 results in the following estimation: $v_{cm} \cong \frac{1}{2}\frac{\gamma}{\eta}\Delta\alpha \cong 0.5 \frac{m}{s}$. The value of the change in the interfacial angle, inspired by the magnetic field, namely $\Delta\alpha = 1.0 \pm 0.2^0$ was established experimentally (see the Supplementary Material). It is recognized from the rough estimation, arising from Eq. 5 that such a small change in the interfacial angle may supply to the floating raft relatively high velocities. The estimation supplied by

Eq.6 for the velocity of the steady motion of the center mass of the raft is obviously strongly exaggerated. Actually, the contact angle hysteresis decreased markedly in a course of raft's motion, as observed in the experiment.

We conclude that it is possible to displace a floating millimetrically-scaled polymer raft with the magnetic field of *ca*. 0.06 T. The effect is due to the magnetically induced deformation of the water/vapor interface resulting in the interfacial contact angle hysteresis, eventually driving the raft. The similar effect of displacement of floating liquid marbles [25, 28] with the steady magnetic field of 50 mT was reported recently in Ref. 24. The reported effect may be effectively exploited for the remote transport of chemical and biological mini-cargos. Note, that the effects due to the elasto-capillarity [29], which may be important for floating of deformable polymer rafts, have been neglected within our analysis.

**Acknowledgements**

We are indebted to Mrs. Yelena Bormashenko for her inestimable help in preparing this manuscript. Acknowledgement is made to the donors of the Israel Ministry of Absorption for the partial support of the scientific activity of Dr. Mark Frenkel.

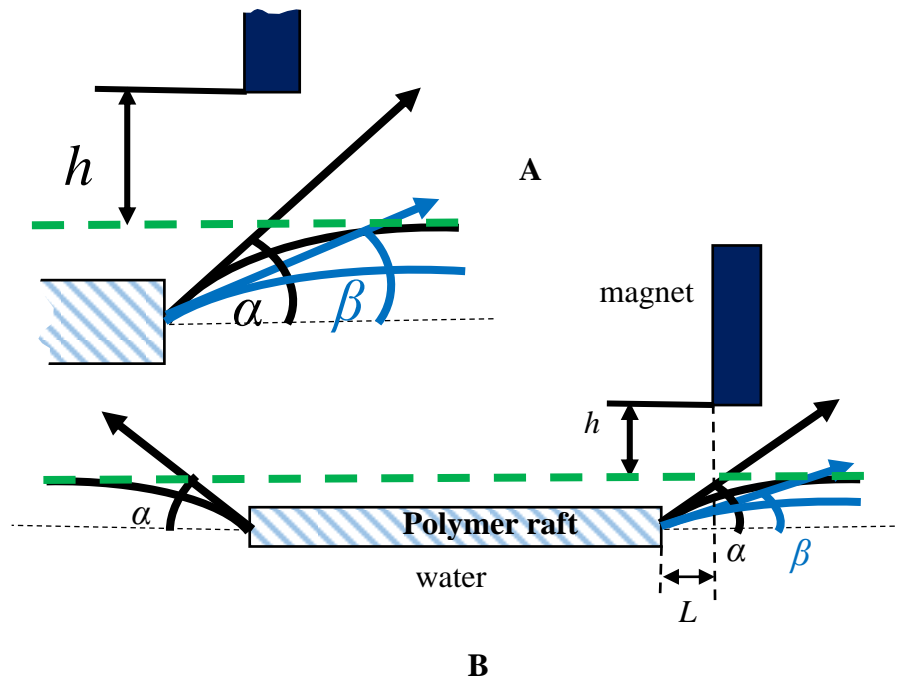

**Figure 1**. Polymer raft floating on the water surface is depicted; $\alpha$ and $\beta$ are the interfacial contact angles. The green dashed line shows the water level far from the raft. The interfacial contact angle $\alpha$ corresponds to the zero magnetic field; the interfacial contact angle $\beta<\alpha$ arises from the deformation of the water/vapor interface by the permanent magnet, creating the magnetic field of $B \sim 0.1\text{T}$. **A**. The enlarged area of the water/vapor interface close to the floating raft is depicted. **B**. Side view of the floating raft.

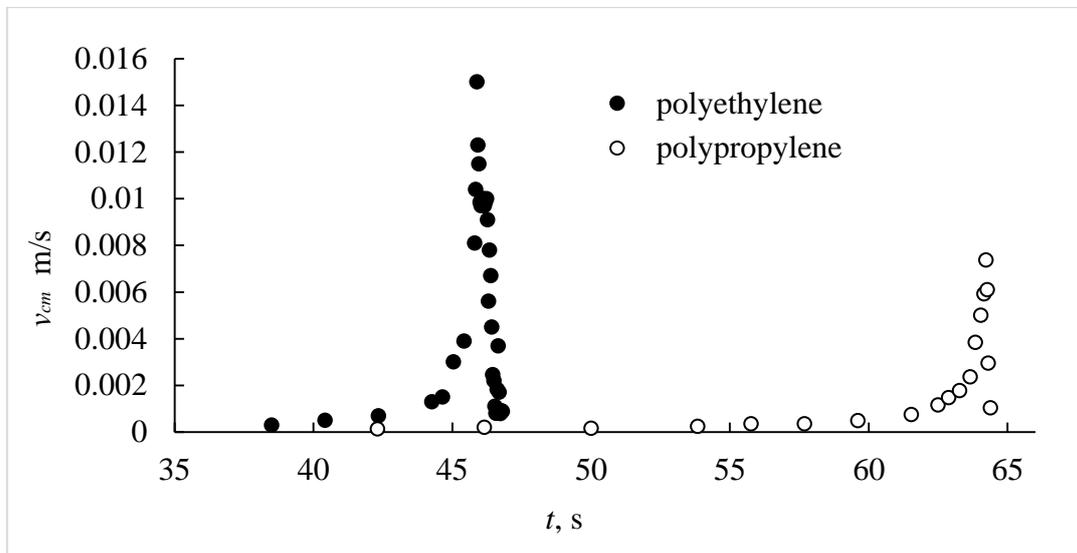

**Figure** 2. The time dependence of the velocity of the center of mass of the polymer rafts $v_{cm}$ is depicted. Black circles correspond to the PE raft. Open circles correspond to the PP raft. The initial horizontal distance of rafts measured from vertical axis of symmetry of the magnet was 20 mm. The motion stopped when rafts came to the magnet, as shown in **Figure 4**.

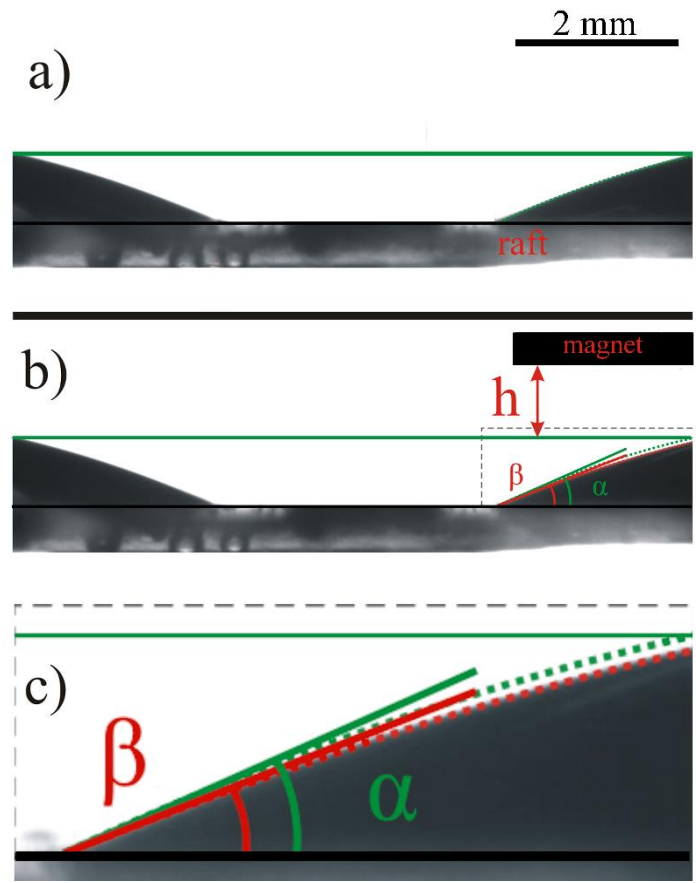

**Figure 3**. Deformation of the water/vapor interface in the vicinity of the floating PE raft induced by the permanent magnet is shown. a) The non-deformed interface is presented; b) the same interface deformed by the magnetic field of *ca*. 0.1 T is depicted. c) The magnified fragment of the water/vapor interface deformed by the magnetic field is shown.

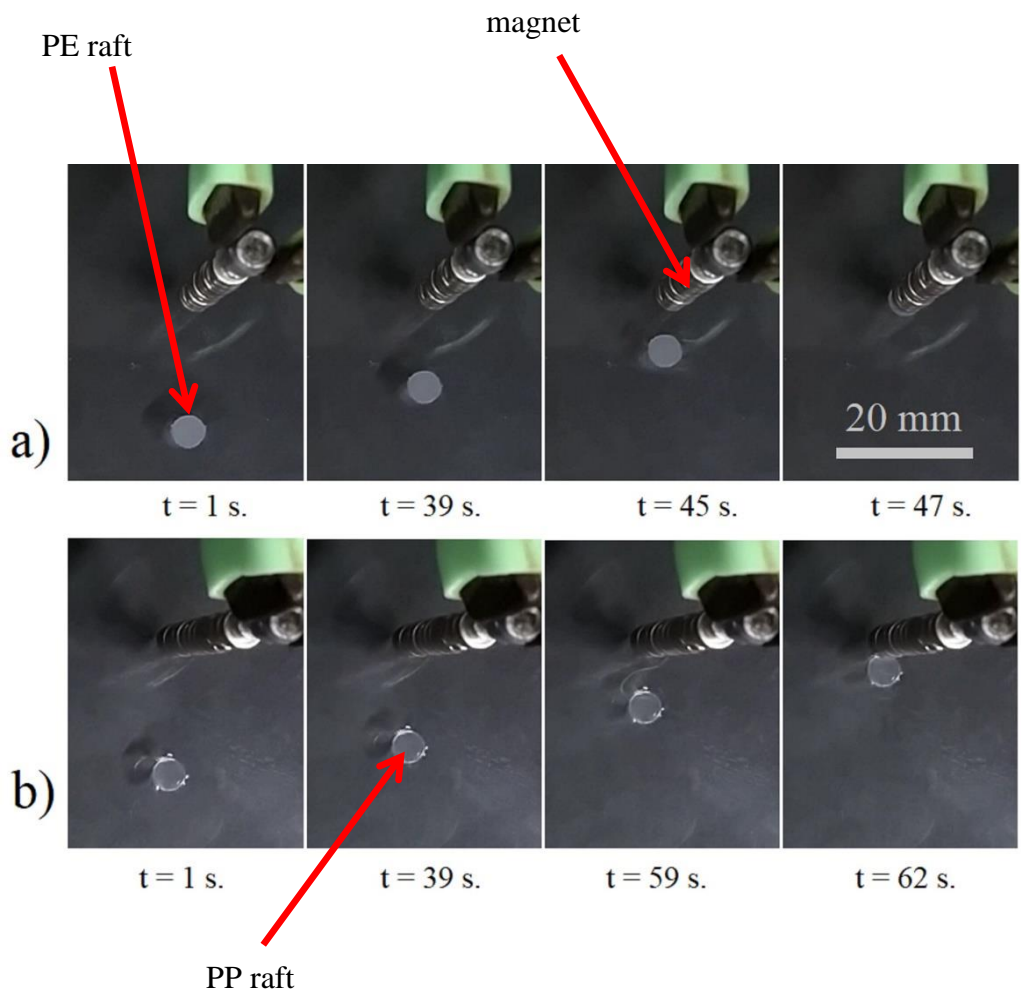

**Figure 4**. The sequence of images, demonstrating the movement of PE (a) and PP (b) rafts towards the magnet. PP is less hydrophobic than PE and the contact angle hysteresis creates the equilibrium location at the certain distance from the vertical axis of the magnet.

**TOC Image**

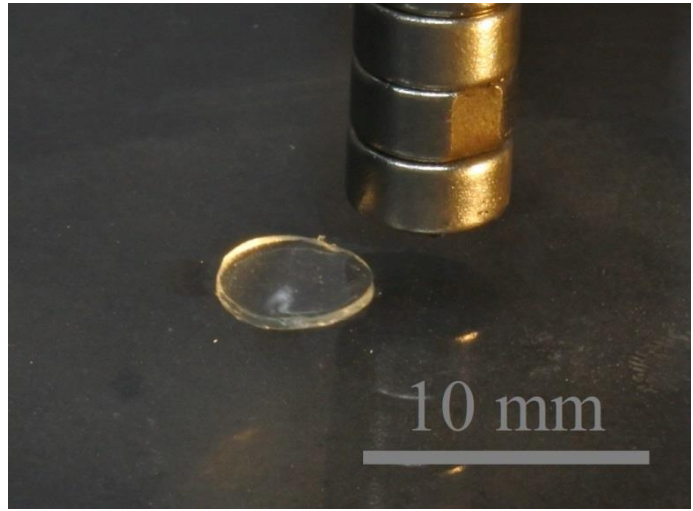